\documentclass[twocolumn,3p,times]{elsarticle}

\usepackage{lineno,hyperref,pifont,natbib,geometry,fleqn,graphicx,txfonts,amsmath,amsthm,amssymb,amsbsy}\usepackage{subfigure,amsfonts,graphicx}
\modulolinenumbers[5]

\journal{Journal of \LaTeX\ Templates}
\graphicspath{{figure/}}

\newcommand{\bra}[1]{\langle#1|}
\newcommand{\ket}[1]{|#1\rangle}

\providecommand{\openone}{\leavevmode\hbox{\small1\kern-3.8pt\normalsize1}}

\begin{document}

\begin{frontmatter}

\title{Coherence and entanglement dynamics of vibrating qubits}

\author[rvt]{Ali Mortezapour\corref{cor1}}
\ead{mortezapour@guilan.ac.ir}
\author[focal]{Ghasem Naeimi}
\author[els,extra]{Rosario Lo Franco\corref{cor1}}
\ead{rosario.lofranco@unipa.it}
\cortext[cor1]{Corresponding author}
\address[rvt]{Department of Physics, University of Guilan, P. O. Box 41335-1914, Rasht, Iran}
\address[focal]{Physics Groups, Qazvin Branch, Islamic Azad University, Qazvin, Iran}
\address[els]{Dipartimento di Energia, Ingegneria dell’Informazione e Modelli Matematici, Università di Palermo, Viale delle Scienze, Edificio 9, 90128 Palermo, Italy}
\address[extra]{Dipartimento di Fisica e Chimica, Università di Palermo, via Archirafi 36, 90123, Palermo, Italy}

\begin{abstract}
We investigate the dynamics of coherence and entanglement of vibrating qubits. Firstly, we consider a single trapped ion qubit inside a perfect cavity and successively we use it to construct a bipartite system made of two of such subsystems, taken identical and noninteracting. As a general result, we find that qubit vibration can lead to prolonging initial coherence in both single-qubit and two-qubit system. However, despite of this coherence preservation, we show that the decay of the entanglement between the two qubits is sped up by the vibrational motion of the qubits. Furthermore, we highlight how the dynamics of photon-phonon correlations between cavity mode and vibrational mode, which may serve as a further useful resource stored in the single-qubit system, is strongly affected by the initial state of the qubit. These results provide new insights about the ability of systems made of moving qubits in maintaining quantum resources compared to systems of stationary qubits.
\end{abstract}


\end{frontmatter}


\section{Introduction}
Quantum coherence and entanglement are the two most significant features of quantum theory which emerge due to the superposition principle \cite{RevModPhys.89.041003,horodecki2009quantum,LFCSciRep,lofrancoreview,Orszag:10}. Generally, a system consisting of two or more subsystems is said to be entangled if its quantum state cannot be described as a simple product of the quantum states of the constituent subsystems, which means that the state is not separable. Nowadays, it has been recognized that quantum entanglement is an essential tool for quantum information processes, such as quantum teleportation \cite{PRL701895}, quantum error correction \cite{PRL77198,PRA5567}, quantum cryptography \cite{PRL67661} and quantum dense coding \cite{PRL692881}. 
On the other hand, quantum coherence is more fundamental than entanglement. Quantum coherence not only exists in multipartite systems but also in single-partite systems. Recent studies suggest that quantum coherence can be employed as a resource, similarly to entanglement, in various quantum information tasks \cite{RevModPhys.89.041003,arXiv0612146,PRL113140401,PRL116120404,PRL117020402,PhysRevLett.117.160402}. Several proposals have been recently put forward to define valid measures for quantifying coherence in quantum systems \cite{RevModPhys.89.041003,arXiv0612146,PRA94052336,PRL116150502,PRA93012110,PRA94060302}.

Atom-photon interactions provide a convenient way to generate and manipulate quantum coherence and entanglement. The simplest situation of atom-photon interaction is that of a two-level atom inside a cavity sustaining a single electromagnetic field mode, described by the famous Jaynes-Cummings Hamiltonian \cite{IEEE}. In this context, it is well known that the coupling of an atom to the cavity field is position-dependent, which in turn makes the atom-field coupling for a moving atom qubit time dependent \cite{LPL14055201,OpenSyst241740006}. Recent developments in cavity quantum electrodynamics (QED) setups offer the possibility to trap an ion inside a cavity \cite{RevModPhys.82.1209,BlattReview}. In a Paul trap system, the trapping potential can be approximated to be harmonic. Hence, the center-of-mass motion of an ion in such a trap behaves as a standard harmonic oscillator. It was shown that, in the Lamb-Dicke regime, the quantized harmonic center-of-mass motion of a single two-level ion, similar to the Jaynes-Cummings model (JCM), can be coupled to its own internal electronic states while the ion is interacting with a classical single-mode travelling field \cite{EurLet17509}. This model has been then utilized for a single two-level trapped ion inside a single-mode high-Q cavity \cite{PRA562352}, demonstrating that the interaction of a cavity quantized mode with the trapped ion, within the Lamb-Dicke approximation, can lead to the generation of Greenberger-Horne-Zeilinger states. 
Owing to the analogy between an ion vibrating in a trapping potential and an ion interacting with a quantized cavity field, many effects and ideas observed in the context of cavity QED, such as quantum state engineering \cite{PRL761055,PRA592920,EPJD54715,lofrancoPLA,lofranco2006PRA,lofranco2007PRA,lofranco2006OpenSys}, quantum computing \cite{PRL744091,PhyRep469155} and quantum state endoscopy \cite{PRA514963,PRA532736} can be extended to the trapped ion models. Moreover, the physics of trapped ions has allowed researchers to propose some schemes for generating entanglement. For instance, a scheme for generating a phonon-photon Bell-type state has been introduced \cite{PRA64024305} and the time behavior of entanglement between cavity mode and vibrational mode (mode-mode entanglement) for a system of two trapped ions inside a leaky cavity has been investigated \cite{PRA71063817}. A further study has been carried out concerning mode-mode entanglement in a system made of a cluster of $N$ trapped ions interacting dispersively with a quantized electromagnetic field \cite{PRA75042315}.

In spite of these studies, the evolution of coherence and mode-mode correlations of a single trapped ion qubit has not been investigated so far. Furthermore, entanglement and coherence dynamics of independent trapped vibrating qubits inside  separated cavities has remained unexplored. Since separated qubit subsystems represent one of the preferred scenarios for quantum networks, the evolution of such systems deserves a dedicated case study.
Motivated by these considerations, we first focus on a system containing a trapped single two-level ion (qubit) inside a high-Q cavity which is interacting with its vibrational degrees of freedom and cavity modes. Subsequently, we extend our analysis to a system composed of two of such subsystems, which are separated and initially entangled. We strive to comprehend how the center-of-mass motion of the qubits influences the dynamics of coherence and entanglement. We also examine the effect of qubit initial state, intensity of both cavity and vibrational modes on the mode-mode correlation.

The paper is organized as follows. In Sec.~\ref{sec2}, we present the results about the effect of qubit, cavity and vibrational parameters on the dynamics of ion qubit coherence and mode-mode correlation. In Sec.~\ref{sec3}, we extend the study to the bipartite system, analyzing the evolution of entanglement and coherence between two separated identical trapped qubits for different values of the parameters. In Section.~\ref{sec4} we give our conclusions.

\section{Single-qubit system}\label{sec2}

The system under investigation is a single two-level ion (qubit) trapped in a linear Paul trap and located inside a single-mode high-Q cavity. Owing to the confinement of the qubit in the Paul trap, the qubit vibrates with a high frequency comparable to or larger than the fundamental frequency of the cavity field. We assume that the trap axis coincides with the axis of the cavity so that, as already discussed \cite{PRA562352,PRA64024305}, the internal states of the qubit (namely, the excited state $\lvert e\rangle$ and the ground state $\lvert g\rangle$) are coupled to both the cavity field (cavity mode) and the vibrational degrees of freedom (vibrational mode). The Hamiltonian corresponding to such a system is given by
\setlength\arraycolsep{1.4pt}\begin{eqnarray}
\label{eq:1}
\hat{H}&=&\hat{H}_{0} +\hat{H}_{int},\nonumber\\
\hat{H}_{0}&=& \frac{\hbar\omega_{0}}{2}\sigma_{z}+\hbar\omega_\mathrm{v}\left(\hat{a}^{\dagger}\hat{a}+\frac{1}{2}\right)+\hbar\omega\left(\hat{b}^{\dagger}\hat{b}+\frac{1}{2}\right), \nonumber \\
\hat{H}_{int}&=& \hbar \kappa \sin[\eta(\hat{b}+\hat{b}^{\dagger})](\hat{\sigma}_{+}+\hat{\sigma}_{-})(\hat{a}+\hat{a}^{\dagger}),
\end{eqnarray} 
where $\hat{a}^{\dagger}(\hat{a})$ is the creation (annihilation) operator for the cavity mode with frequency $\omega$, $\hat{b}^{\dagger}(\hat{b})$ denotes the creation (annihilation) operator of the center-of-mass vibrational motion of the qubit with frequency $\omega_\mathrm{v}$, $\hat{\sigma}_{+}=\lvert e\rangle \langle g\rvert$, $\hat{\sigma}_{-}=\lvert g\rangle \langle e\rvert$ and $\hat{\sigma}_{z}=\lvert e\rangle \langle e\rvert-\lvert g\rangle \langle g\rvert$ are the qubit operators, $\kappa$ is the coupling constant between cavity mode and qubit. Moreover, $\eta$ represents the Lamb-Dicke parameter.

We suppose the trapped qubit is constrained in the Lamb-Dicke regime, and the Lamb-Dicke parameter meets the condition $\eta\ll1$. In this regime, $\hat{H}_{int}$ can be approximated by the expansion to the first order in $\eta$ as
\begin{equation} 
\label{eq:2}  
\hat{H}_{int}=\hbar \kappa \eta (\hat{a}+\hat{a}^{\dagger})(\hat{\sigma}_{+}+\hat{\sigma}_{-})(\hat{b}+\hat{b}^{\dagger}).
\end{equation}
In the following, we investigate a special case in which the cavity field is tuned to the first red sideband: $\omega_{0}-\omega=\omega_\mathrm{v}$. Under this condition, by dropping the rapidly oscillating terms, the interaction picture Hamiltonian becomes
\begin{equation} 
\label{eq:3}  
\hat{H}_{I}^{r}=\hbar \kappa \eta(\hat{\sigma}_{+}\hat a\hat b+\hat{\sigma}_{-}\hat{a}^{\dagger}\hat{b}^{\dagger}).
\end{equation}
Let us take the system initially in a product state with the ion qubit in a coherent superposition of its internal states $\ket{\psi}=C_{e}\lvert e\rangle+C_{g}\lvert g\rangle$ ($\left|C_{e} \right|^{2}+\left|C_{g} \right|^{2}=1$) while the qubit center-of-mass motion and the cavity field are, respectively, in the coherent states $\ket{\alpha}=\sum_{m}w_{m}\lvert m\rangle $ and $\ket{\beta}=\sum_{n}w_{n}\lvert n\rangle$, where $\lvert m \rangle$, $\lvert n \rangle$ are the excitation number (Fock) states while $w_{m}=e^{-\lvert\alpha\rvert^{2}/2}\lvert\alpha\rvert^{m}/\sqrt{m!}$ and $w_{n}=e^{-\lvert\beta\rvert^{2}/2}\lvert\beta\rvert^{n}/\sqrt{n!}$ denote the coherent distribution of the number states: $\lvert\alpha\rvert^{2}$ is the mean value of the photon number in the cavity field and $\lvert\beta\rvert^{2}$ is the mean value of the phonon number. The overall initial state is thus
\begin{equation} 
\label{eq:4}  
\lvert\Psi_\mathrm{tot}(0)\rangle=\sum_{m}w_{m}\lvert m\rangle\otimes\sum_{n}w_{n}\lvert n\rangle\otimes
(C_{e}\lvert e\rangle+C_{g}\lvert g\rangle),
\end{equation} 
so that, at any later time, the state vector of the system can be written as
\setlength\arraycolsep{1.4pt}\begin{eqnarray}
\label{eq:5} 
\lvert\Psi_\mathrm{tot}(t)\rangle &=& \sum_{m,n}\{[C_{e}A_{m,n}(t)+C_{g}B_{m,n}(t)]\lvert m \rangle\lvert n\rangle\lvert e\rangle \nonumber \\
&&+[C_{g}C_{m,n}(t)+C_{e}D_{m,n}(t)]\lvert m\rangle\lvert n\rangle\lvert g\rangle\}
\end{eqnarray} 
The time-dependent coefficients $A_{m,n}(t)$, $B_{m,n}(t)$, $C_{m,n}(t)$ and $D_{m,n}(t)$ can be found by substituting Eq.~(\ref{eq:5}) into the Schr\"{o}dinger equation $i\hbar\partial\lvert\Psi_\mathrm{tot}(t)\rangle/\partial t=\hat H_{I}^{r}\lvert\Psi_\mathrm{tot}(t)\rangle$ that gives
\begin{eqnarray}
\label{eq:6} 
A_{m,n}(t)&=& w_{m}w_{n}\cos{[\eta \kappa t\sqrt{(m+1)(n+1)}]}, \nonumber \\
B_{m,n}(t)&=&-iw_{m+1}w_{n+1}\sin{[\eta \kappa t\sqrt{(m+1)(n+1)}]}, \nonumber \\
C_{m,n}(t)&=& w_{m}w_{n}\cos{[\eta \kappa t\sqrt{mn}]}, \nonumber \\
D_{m,n}(t)&=&-iw_{m}w_{n}\sin{[\eta \kappa t\sqrt{mn}]}.
\end{eqnarray} 
Taking the partial trace of the global density matrix $\rho_\mathrm{tot}(t)=\lvert\Psi_\mathrm{tot}(t)\rangle\langle\Psi_\mathrm{tot}(t)\lvert$ over the cavity field and vibrational mode degrees of freedom, the reduced density matrix of the qubit in the basis $\{\ket{e},\ket{g}\}$ results to be
\begin{equation} 
\label{eq:8}  
\rho_\mathrm{q}(t)=\left(\begin{array}{cc} \rho_{ee}(t) &  \rho_{eg}(t) \\ 
 \rho_{ge}(t) &  \rho_{gg}(t)\\ 
\end{array}
\right), 
\end{equation}
where
\begin{eqnarray}
\label{eq:9} 
\rho_{ee}(t) &=&\sum_{m,n}\lvert C_{e}A_{m,n}(t)+C_{g}B_{m,n}(t)\rvert^{2},\nonumber \\
\rho_{gg}(t) &=&\sum_{m,n}\lvert C_{g}C_{m,n}(t)+C_{e}D_{m,n}(t)\rvert^{2}=1-\rho_{ee}(t), \nonumber \\
\rho_{eg}(t) &=&\sum_{m,n}[(C_{e}A_{m,n}(t)+C_{g}B_{m,n}(t))\nonumber\\
&&\times (C_{g}C_{m,n}(t)+C_{e}D_{m,n}(t))^{*}]=\rho_{ge}^\ast(t).
\end{eqnarray}

\subsection{Coherence dynamics of the qubit}
We now study the effect of cavity and vibrational parameters on the time evolution of coherence in our qubit system. Many bona-fide quantifiers of quantum coherence have been introduced \cite{RevModPhys.89.041003}.
Among these quantifiers, we adopt an intuitive measure which relies on the off-diagonal elements of the target quantum state which is defined by \cite{PRL113140401}
\begin{equation} 
\label{eq:10} 
\zeta(t)=\sum_{i,j\ (i\neq j)}\lvert\rho_{ij}(t)\rvert,
\end{equation}
where $\rho_{ij}(t)$ $(i\neq j)$ are the off-diagonal elements of the system density matrix $\rho (t)$. 

\begin{figure}[t!]
\includegraphics[scale=.44]{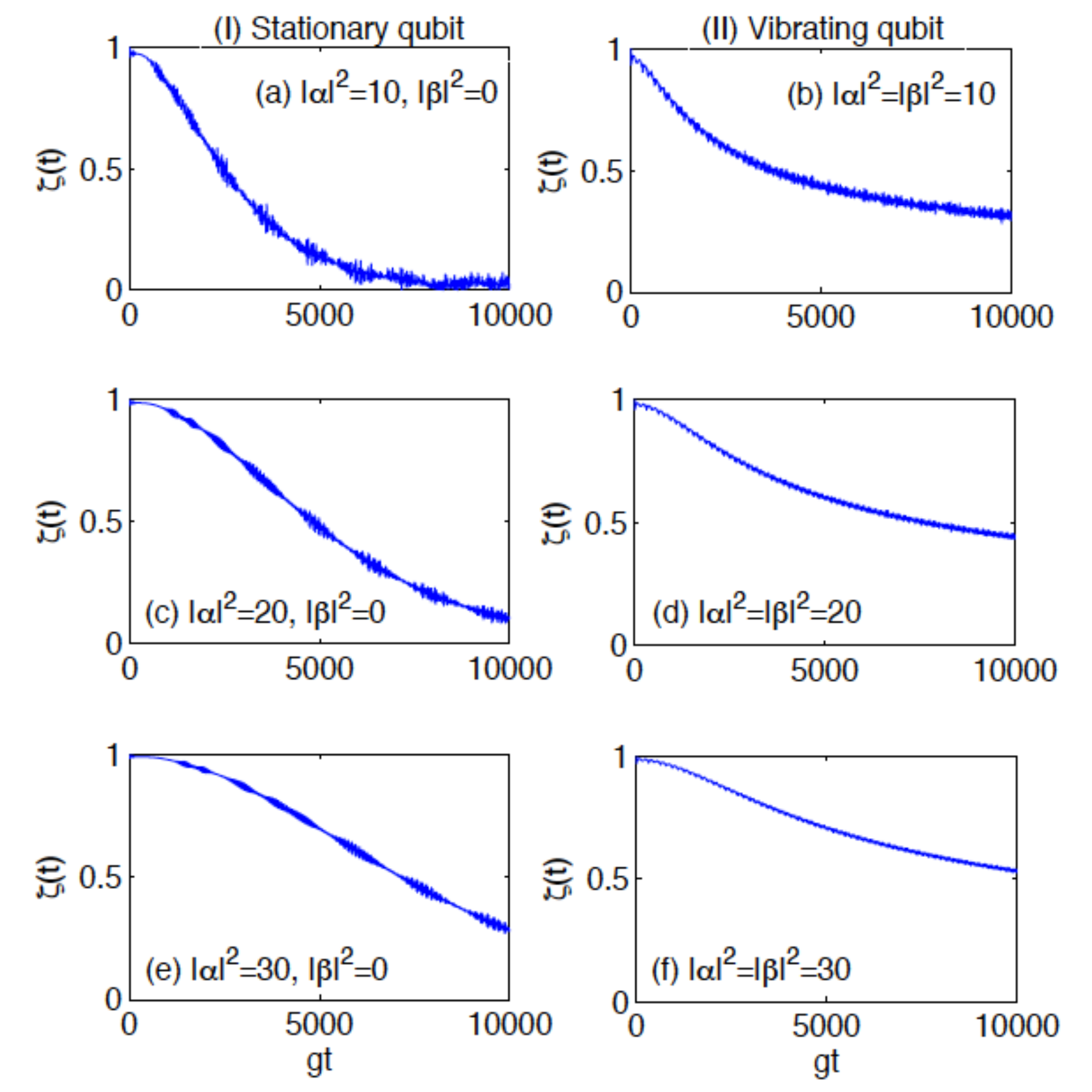}
\caption{The dynamic behavior of coherence of stationary qubit (column I) and vibrating qubit (column II) for different intensities of cavity and vibrational modes. The other parameters are $C_{e}=C_{g}=1/\sqrt{2}$ ($\zeta(0)=1$), $\eta=0.02$, $\kappa=1$.}
\label{opt1}
\end{figure}

We easily obtain the single-qubit coherence evolution by using $\zeta(t)$ with the density matrix $\rho_\mathrm{q}(t)$ of Eq.~(\ref{eq:8}). Fig.~\ref{opt1} illustrates the effect of intensity of cavity and vibrational modes on the time evolution of coherence starting from a maximally coherent state ($\zeta(0)=1$) in the basis $\{\ket{e},\ket{g}\}$, for a stationary qubit (column I) and vibrating qubit (column II). As can be seen when the qubit is motionless, increasing the intensity of the cavity field (larger mean photon number) preserves the initial coherence for longer times. However, coherence preservation is more effective for vibrating qubits. In fact, as displayed in the plots of column II of Fig.~\ref{opt1}, increasing the intensity of vibrational modes (larger mean phonon number) guarantees better coherence preservation. 

\begin{figure}[t!]
\includegraphics[scale=.44]{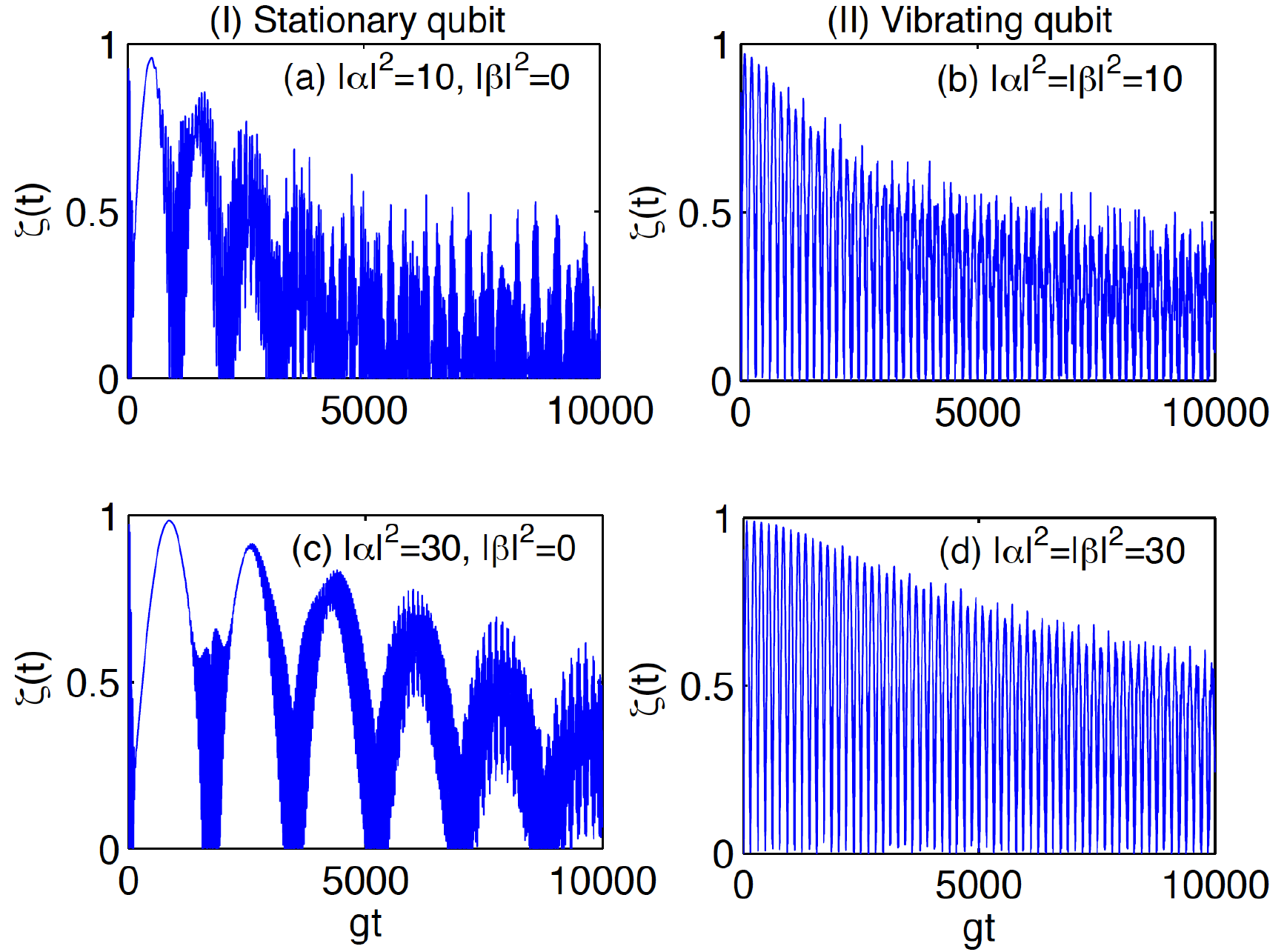}
\caption{The dynamic behavior of coherence of vibrating qubit for different intensities of vibrational and cavity  modes. The qubit is initially prepared in its excited state ($C_{e}=1, C_{g}=0$ ) or $\zeta(0)=0$. The other parameters are taken as in Fig.~\ref{opt1}.}
\label{opt2}
\end{figure}

In Fig.~\ref{opt2}, we divert our attention to the case when initial coherence of the qubit in the basis $\{\ket{e},\ket{g}\}$ is zero, making the choice $C_{e}=1$ (qubit initially in the excited state). In column I of Fig.~\ref{opt2}, one can see that interaction of cavity field with the stationary qubit gives rise to an oscillating coherence with gradual reduction in the maximum value of oscillation (envelope) as time goes by. The plots of the second column, corresponding to the vibrating qubit, show an oscillating behavior for the coherence with frequency significantly larger than the case of stationary qubit. Furthermore, increasing the intensity of both cavity and vibrational modes lead to an increase of the maximum values of the coherence. These observations confirm that the qubit coherence generated during the evolution decays more slowly for a vibrating qubit. 

\subsection{Two-mode correlation}
We now investigate the existence of correlations between vibrational and cavity modes (mode-mode correlation), which may serve as a further quantum resource within the single-qubit system.
Correlations between any couple of modes can be obtained by calculating the intermode second-order coherence \cite{PRA458037,RepProgPhys4358,PRA86023810,ExpThePhy122984}, which in our system is given by
\begin{eqnarray}
\label{eq:11} 
G_{ab}^{2}(t)&=&\frac{\langle\hat a^{\dagger}(t)\hat b^{\dagger}(t)\hat b(t)\hat a(t)\rangle}{\langle\hat a^{\dagger}(t)\hat a(t)\rangle \langle\hat b^{\dagger}(t)\hat b(t)\rangle}\nonumber \\
&=& 1+\frac{C(t)}{\langle \hat a^{\dagger}(t)\hat a(t)\rangle \langle \hat b^{\dagger}(t)\hat b(t)\rangle},
\end{eqnarray} 
where the cross-correlation function
\begin{equation}
\label{eq:13} 
C(t)=\langle \hat a^{\dagger}(t)\hat b^{\dagger}(t)\hat b(t)\hat a(t)\rangle -
\langle \hat a^{\dagger}(t)\hat a(t)\rangle \langle \hat b^{\dagger}(t)\hat b(t)\rangle, 
\end{equation}
characterizes the correlation between cavity and vibrational modes. In our quantitative estimation, we thus use the cross-correlation function $C(t)$ instead of $G_{ab}^{2}$. According to this measure, the two modes are correlated whenever $C(t)>0$, whereas they will be anti-correlated if $C(t)<0$. Furthermore, for $C(t)=0$ the two modes are uncorrelated and independent. The average values appearing in Eq.~(\ref{eq:13}) are calculated on the reduced state after tracing out the suitable degrees of freedom from the global density matrix $\rho_\mathrm{tot}(t)=\lvert\Psi_\mathrm{tot}(t)\rangle\langle\Psi_\mathrm{tot}(t)\lvert$, with $\lvert\Psi_\mathrm{tot}(t)\rangle$ given in Eq.~(\ref{eq:5}). 

\begin{figure}[t!]
\includegraphics[scale=.44]{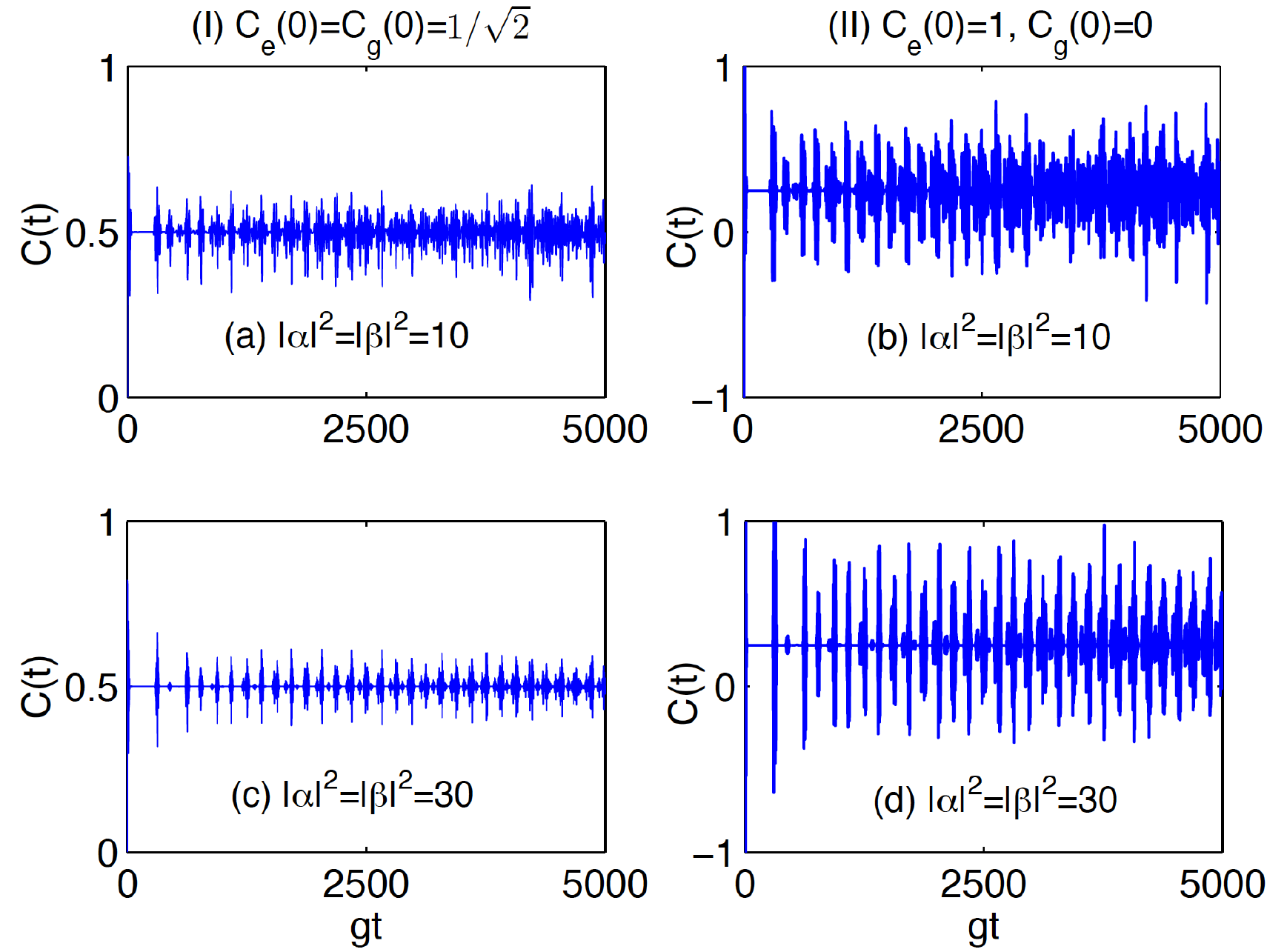}
\caption{The dynamics of mode-mode correlation for an initial maximally coherent qubit $C_{e}=C_{g}=1/\sqrt{2}$ (column I) and for an initially incoherent qubit $C_{e}=1$, $C_{g}=0$ (column II). The other parameters are the same of Fig.~\ref{opt1}.}
\label{opt3}
\end{figure}

Fig.~\ref{opt3} exhibits the time evolution of the mode-mode correlation $C(t)$. The plots of the first column correspond to the qubit initially prepared in the maximally coherent state $(\ket{e}+\ket{g})/\sqrt{2}$ whereby, after interaction, one interestingly finds that cavity and vibrational modes become permanently correlated. Differently, when the qubit is initially prepared in its excited (incoherent) state, such a behavior does not occur. Moreover, the system undergoes changes in the fluctuation amplitudes of $C(t)$ as a result of increasing intensities of cavity and vibrational modes.

\section{Two-qubit system }\label{sec3}
In this section we study a composite system made of two separated, identical and noninteracting subsystems, labeled as 1 and 2, respectively. Each one of these subsystems is assumed to be the same as that considered in the previous section, that is a qubit which can vibrate inside a single-mode cavity. The two subsystems, being identical, will have the same intensities of cavity and vibrational modes (mean photon and phonon numbers), that is $|\alpha_1|=|\alpha_2|=|\alpha|$ and $|\beta_1|=|\beta_2|=|\beta|$. 

The two qubit are initially prepared in the Bell-like states
\begin{equation}
\label{eq:16}
\lvert\Phi\rangle = \mu\lvert e_{1}g_{2}\rangle+\upsilon\lvert g_{1}e_{2}\rangle, \
\lvert\Psi\rangle = \mu\lvert e_{1}e_{2}\rangle+\upsilon\lvert g_{1}g_{2}\rangle,
\end{equation}
with $|\mu|^2+|\upsilon|^2=1$ and $\lvert e_{1}g_{2}\rangle \equiv \ket{e_1}\ket{g_2}=\ket{e_1}\otimes\ket{g_2}$ (analogously for the other two-qubit terms).
To describe the composite system, we choose the standard (computational) product basis $\{\lvert 1\rangle\equiv\lvert e_{1}e_{2}\rangle, \lvert 2\rangle\equiv\lvert e_{1}g_{2}\rangle, \lvert 3\rangle\equiv\lvert g_{1}e_{2}\rangle, \lvert 4\rangle\equiv\lvert g_{1}g_{2}\rangle\}$.                   
From the reduced density matrix of Eq.~(\ref{eq:8}) and due to the identity of the two subsystems, it is straightforward to see that a time evolution operator and its adjoint for the single qubit $i$ ($i=1, 2$) in the basis $\{\lvert e_{i}\rangle$, 
$\lvert g_{i}\rangle\}$ can be suitably introduced as
\begin{eqnarray}
\label{eq:17}
\hat{U}_{i}^{r}(t)&=&\sum_{m,n}\big(A_{m,n}^{i}(t)\lvert e_{i}\rangle\langle e_{i}\rvert + C_{m,n}^{i}(t)\lvert g_{i}\rangle\langle g_{i}\rvert \nonumber \\
&+& B_{m,n}^{i}(t)\lvert e_{i}\rangle\langle g_{i}\rvert + D_{m,n}^{i}(t)\lvert g_{i}\rangle\langle e_{i}\rvert\big),\nonumber\\
\hat{U}_{i}^{r\dagger}(t)&=&\sum_{m,n}\big(A_{m,n}^{i\ast}(t)\lvert e_{i}\rangle\langle e_{i}\rvert + C_{m,n}^{i\ast}(t)\lvert g_{i}\rangle\langle g_{i}\rvert \nonumber \\
&+& B_{m,n}^{i\ast}(t)\lvert g_{i}\rangle\langle e_{i}\rvert + D_{m,n}^{i\ast}(t)\lvert e_{i}\rangle\langle g_{i}\rvert\big).
\end{eqnarray}
These operators are such that, given an arbitrary initial pure state of the single qubit $\rho_\mathrm{q}^i(0)=\ket{\psi^i}\bra{\psi^i}$ with $\ket{\psi^i}=C_{e}^i\lvert e\rangle+C_{g}^i\lvert g\rangle$, the single-qubit evolved density matrix of Eq.~(\ref{eq:8}) is retrieved by $\rho_\mathrm{q}^i(t)=\hat U_{i}^{r}(t)\rho_\mathrm{q}^i(0)\hat U_{i}^{r\dagger}(t)$.
 
Therefore, exploiting the evolution operators of Eq.~(\ref{eq:17}) and the independence of the qubits \cite{lofrancoreview,BLFC2007PRL}, the evolved density matrices of the composite system, corresponding to the two initial states of Eq.~(\ref{eq:16}), can be respectively obtained by
\begin{eqnarray}
\label{eq:18}
\hspace*{-0.5cm}
\rho_{\Phi}(t)&=&\hat U_{1}^{r}(t)\otimes\hat U_{2}^{r}(t)\lvert\Phi\rangle\langle\Phi\rvert\hat U_{1}^{r\dagger}(t)\otimes\hat U_{2}^{r\dagger}(t) \nonumber \\
&=&\lvert\mu\rvert^{2}\hat U_{1}^{r}(t)\lvert e_{1}\rangle\langle e_{1}\rvert\hat U_{1}^{r\dagger}(t)\otimes\hat U_{2}^{r}(t)\lvert g_{2}\rangle\langle g_{2}\rvert\hat U_{2}^{r\dagger}(t) \nonumber \\
&+&\lvert\upsilon\rvert^{2}\hat U_{1}^{r}(t)\lvert g_{1}\rangle\langle g_{1}\rvert\hat U_{1}^{r\dagger}(t)\otimes \hat U_{2}^{r}(t)\lvert e_{2}\rangle\langle e_{2}\rvert\hat U_{2}^{r\dagger}(t) \nonumber \\
&+& \mu\upsilon^{*}\hat U_{1}^{r}(t)\lvert e_{1}\rangle\langle g_{1}\rvert\hat U_{1}^{r\dagger}(t)\otimes \hat U_{2}^{r}(t)\lvert g_{2}\rangle\langle e_{2}\rvert\hat U_{2}^{r\dagger}(t)\nonumber \\
&+& \mu^{*}\upsilon\hat U_{1}^{r}(t)\lvert g_{1}\rangle\langle e_{1}\rvert\hat U_{1}^{r\dagger}(t)\otimes \hat U_{2}^{r}(t)\lvert e_{2}\rangle\langle g_{2}\rvert\hat U_{2}^{r\dagger}(t),\nonumber\\
\end{eqnarray}
and
\begin{eqnarray}
\label{eq:19}
\hspace*{-0.5cm}
\rho_{\Psi}(t)&=&\hat U_{1}^{r}(t)\otimes\hat U_{2}^{r}(t)\lvert\Psi\rangle\langle\Psi\rvert\hat U_{1}^{r\dagger}(t)\otimes\hat U_{2}^{r\dagger}(t) \nonumber \\
&=&\lvert\mu\rvert^{2}\hat U_{1}^{r}(t)\lvert e_{1}\rangle\langle e_{1}\rvert\hat U_{1}^{r\dagger}(t)\otimes \hat U_{2}^{r}(t)\lvert e_{2}\rangle\langle e_{2}\rvert\hat U_{2}^{r\dagger}(t)\nonumber \\
&+& \lvert\upsilon\rvert^{2}\hat U_{1}^{r}(t)\lvert g_{1}\rangle\langle g_{1}\rvert\hat U_{1}^{r\dagger}(t)\otimes \hat U_{2}^{r}(t)\lvert g_{2}\rangle\langle g_{2}\rvert\hat U_{2}^{r\dagger}(t)\nonumber \\
&+&\mu\upsilon^{*}\hat U_{1}^{r}(t)\lvert e_{1}\rangle\langle g_{1}\rvert\hat U_{1}^{r\dagger}(t)\otimes \hat U_{2}^{r}(t)\lvert e_{2}\rangle\langle g_{2}\rvert\hat U_{2}^{r\dagger}(t)\nonumber \\
&+& \mu^{*}\upsilon\hat U_{1}^{r}(t)\lvert g_{1}\rangle\langle e_{1}\rvert\hat U_{1}^{r\dagger}(t)\otimes \hat U_{2}^{r}(t)\lvert g_{2}\rangle\langle e_{2}\rvert\hat U_{2}^{r\dagger}(t).\nonumber\\
\end{eqnarray}

\subsection{Entanglement and coherence dynamics}
The entanglement of a two-qubit system, described by the density matrix $\rho$, can be measured by the concurrence which is defined as \cite{PRL785022,PRL802245}
\begin{equation}
\label{eq:20}
\mathcal{C}(\rho)=\mathrm{max}\{0,\sqrt{\lambda_{1}}-\sqrt{\lambda_{2}}-\sqrt{\lambda_{3}}-\sqrt{\lambda_{4}}\},
\end{equation}                              
where the quantities $\lambda_{i}$ are the eigenvalues of the matrix $\rho\tilde{\rho}=\rho(\sigma_{y}\otimes\sigma_{y})\rho^{*}(\sigma_{y}\otimes\sigma_{y})$ which are arranged in decreasing order of magnitude. Here $\rho^{*}$ is the complex conjugate of $\rho$ in the standard basis, and $\sigma_{y}$ denotes the Pauli matrix. The concurrence varies from $0$ for a separable (non-entangled) state to $1$ for a maximally entangled state. 
\begin{figure}[t!]
\includegraphics[scale=.44]{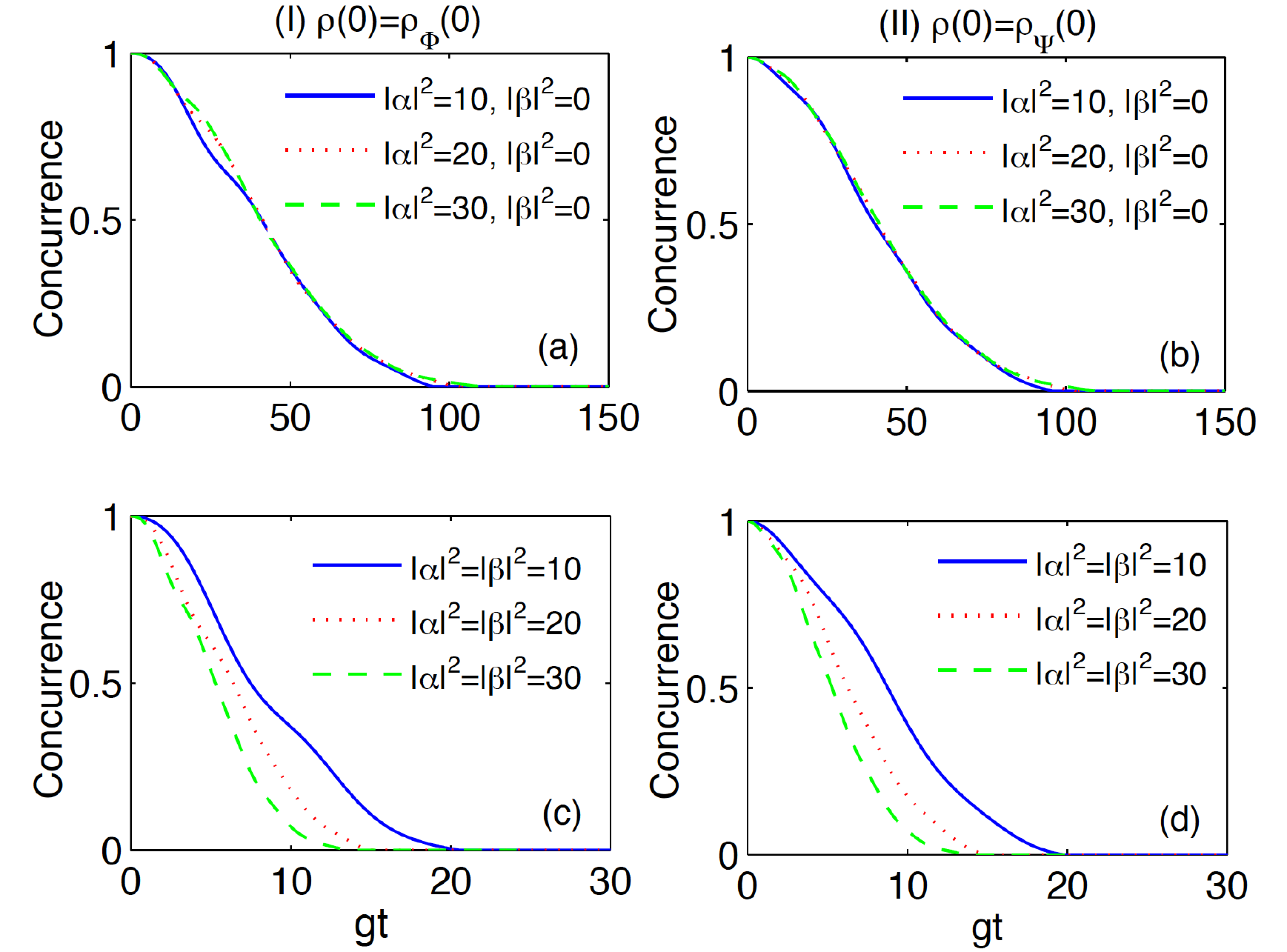}
\caption{The dynamic behavior of concurrence for different intensities of cavity and vibrational modes. Column I for  
$\rho(0)=\rho_{\Phi}(0)$ and column II for $\rho(0)=\rho_{\Psi}(0)$. The other parameters are $\eta_{1}=\eta_{2}=0.02$, $\mu=\upsilon=1/\sqrt{2}$, $\kappa_{1}=\kappa_{2}=\kappa=1$.}
\label{opt4}
\end{figure}
Fig.~\ref{opt4} depicts the effect of the intensities of cavity and vibrational modes of the two qubits on the time evolution of concurrence for two different initial Bell states defined in Eq.~(\ref{eq:16}) for $\mu=\upsilon=1/\sqrt{2}$, namely $\rho_{\Phi}(0)=\ket{\Phi}\bra{\Phi}$ (column I) and $\rho_{\Psi}(0)=\ket{\Psi}\bra{\Psi}$ (column II). We notice that these two initial states are sufficient to acquire the qualitative time behavior of entanglement and coherence. In fact, initial two-qubit states with different values of the probability amplitudes $\mu$, $\upsilon$ are less entangled and will produce only quantitative differences in the plots, leaving the qualitative behavior invariant. 

It is evinced from Fig.~\ref{opt4} that the time behavior of concurrence for the both initial Bell states is approximately the same. As a first qualitative aspect, for stationary qubits it is evident that increasing the intensity of the cavity mode does not appreciably affect the entanglement dynamics (see panels (a) and (b)). Instead, drawing a comparison between panels (a) and (c) or panels (b) and (d) of Fig.~\ref{opt4}, one infers that the entanglement between vibrating qubits disappears earlier compared to the case of stationary qubits. The more the intensity of qubit vibrational modes, the faster the entanglement decays. In other words, the vibration of qubits has a detrimental effect on the entanglement dynamics. This behavior for the two-qubit system contrasts with the result found above for the single-qubit, where the maintenance of coherence is instead enhanced by the presence of an intense vibrational mode for the qubit (large phonon number, see Figs.~\ref{opt1} and \ref{opt2}). This fact highlights how the population decay of a single qubit may qualitatively differ from the entanglement (quantum correlations) decay of a composite system of qubits, as already observed in phenomena such as entanglement sudden death \cite{eberlyScience2007} and entanglement revivals \cite{lofrancoreview}.   

\begin{figure}[t!]
\includegraphics[scale=.28]{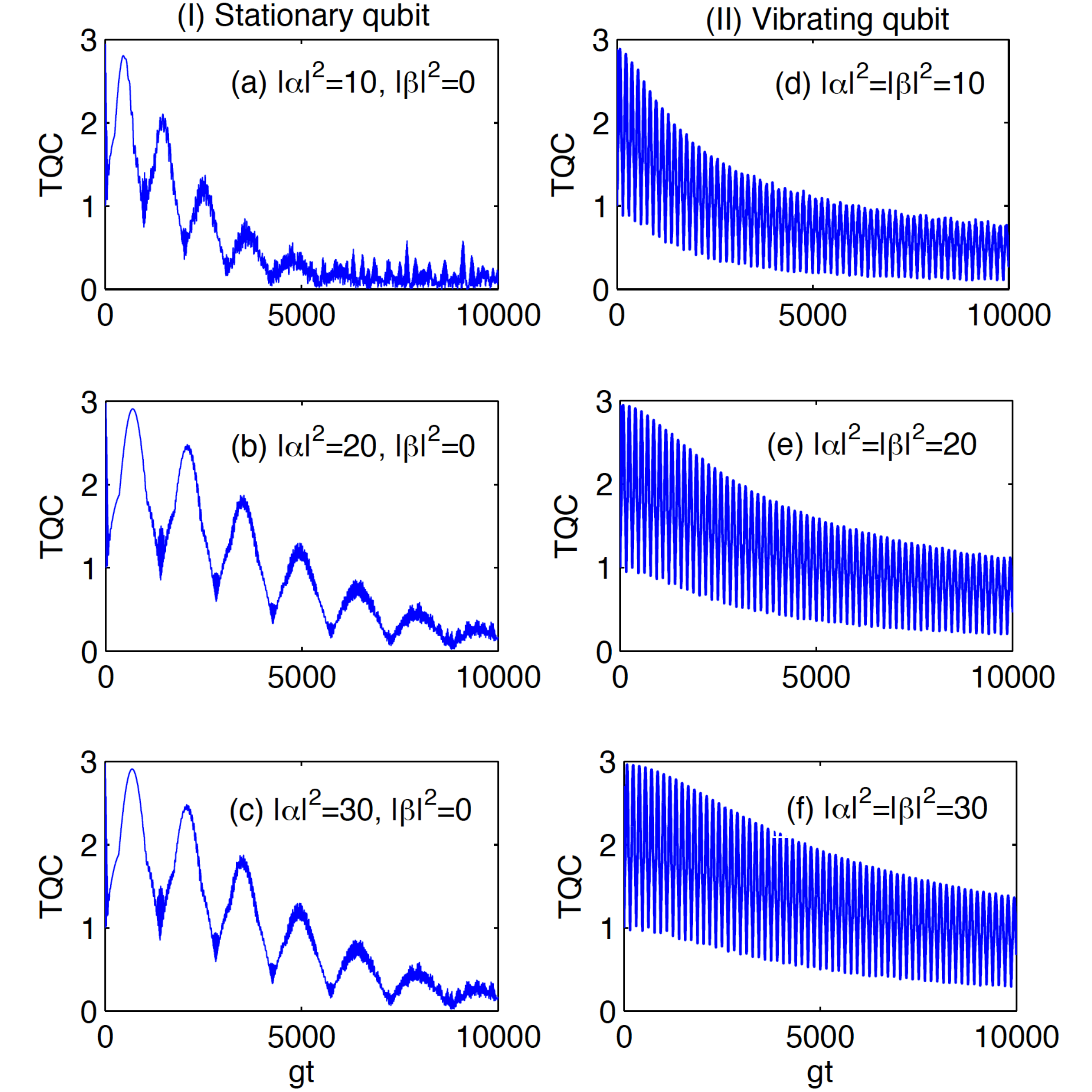}
\caption{The dynamic behavior of two-qubit coherence (TQC) for different intensities of cavity and vibrational modes for $\rho(0)=\rho_{\Phi}(0)$. Column (I) for stationary qubits and column (II) for vibrating qubits. The other parameters are taken as in Fig.~\ref{opt4}.}    
\label{opt5}
\end{figure}

As a final analysis, we apply the definition of coherence in Eq.~(\ref{eq:10}) to the off-diagonal elements of the two-qubit density matrices of Eqs.~(\ref{eq:18}) and (\ref{eq:19}) which supplies the two-qubit coherence (TQC). 
In Fig.~\ref{opt5} we plot the time evolution of TQC for various intensities of cavity and vibrational modes in the case of stationary qubit (column I) and vibrating qubit (column II). The composite system is initially prepared in $\rho_{\Phi}(0)$ (our calculations show that the same results are achieved when $\rho(0)=\rho_{\Psi}(0)$). As can be seen from column I of Fig.~\ref{opt5}, an increase of the intensity of the cavity mode delays the decay of TQC, which oscillates during the evolution. On the other hand, the plots of column II of Fig.~\ref{opt5} display that the vibration of qubits induces fast oscillations in the dynamics of TQC yet prolonging its preservation compared to the case of stationary qubits. The more intense the qubit vibrational mode (that is the larger the mean phonon number), the slower the decay of the two-qubit coherence. This behavior resembles that of the single-qubit coherence obtained in Sec. 2 (see, in particular, Fig.~\ref{opt2}). Recalling the response of the entanglement evolution to qubit vibration discussed above, the results evidence that coherence and entanglement are in general quantum resources of different nature when seen within the same composite system.

\section{Conclusion}\label{sec4}
In this work we have investigated the effect of vibrational degrees of freedom of moving qubits, trapped inside perfect single-mode cavities, on the dynamics of quantum resources such as coherence and entanglement. Here, the cavity losses have not been taken into account in order to allow us to focus just on the role played by the qubit vibration (phonon-photon interaction) in addition to the qubit-cavity interaction. Photon losses will be considered elsewhere to see to which extent  they modify the dynamics of the system. 

Our study has first shown that the presence of vibration of a qubit inside a single-mode cavity is capable to prolong its coherence time with respect to the case when the qubit is at rest and only the cavity mode is activated. In particular, larger mean phonon numbers, which describe a more intense (or energetic) vibration mode, increases the maintenance time of single-qubit coherence. We have also found that such a system can permanently store hybrid correlations between vibrational and cavity modes (phonon-photon correlations), which may serve as a further quantum resource, thanks to the interaction with the qubit. This phenomenon specifically occurs when the qubit is prepared in its maximally coherent state (symmetric superposition of its internal states), thus exhibiting a strong qubit state dependence. This permanent preservation of mode-mode correlations would be jeopardized by photon losses from a leaky cavity, but our result indicates that their detrimental effect can be nevertheless efficiently weakened by adjusting the qubit initial state.  

We have then extended our analysis to two noninteracting qubits in separated cavities, for establishing how the vibrational degrees of freedom of the qubits affect the dynamics of both entanglement and two-qubit coherence. Interestingly, we have discovered that, while the decay of entanglement takes place more rapidly in the presence of vibrational motion of the qubits, the latter instead enhances the preservation of coherence of the two-qubit state, in analogy with the single-qubit coherence. Therefore, depending on which quantum trait of the composite system one desires to exploit, entanglement or coherence, vibrations of the qubits can be a drawback to be avoided or an advantage to be controlled. In conclusion, our work supplies new useful insights about the capability of systems made of moving qubits to protect quantum resources compared to systems of stationary qubits.

\end{document}